\documentclass{gGAF2e} 

\usepackage{graphicx}
\usepackage{dcolumn}
\usepackage{bm}
\usepackage{lineno}
\usepackage{color}
\usepackage{hyperref}

\begin{document}

\title{Dynamical Consequences of Polar Amplification on Standing Rossby Waves: a Laboratory Perspective}

\author{
Kial D. Stewart${^{1,2,3}}$$^{\ast}$\thanks{$^\ast$Corresponding author. Email: kial.stewart@anu.edu.au}, Thomas G. Schmaltz${^{2,4,5}}$ \& Callum J. Shakespeare${^{1,2}}$\\
${^1}$ Climate \& Fluid Physics Laboratory, Australian National University\\
${^2}$ ARC Centre of Excellence for Climate Extremes, University of New South Wales, Australia\\
${^3}$ ARC Centre of Excellence for 21st Century Weather, Monash University, Australia\\
${^4}$ Climate Change Research Centre, University of New South Wales, Australia\\
${^5}$ ARC Centre for Excellence in Antarctic Science, University of Tasmania, Australia}

\maketitle

\linespread{1.5}
\begin{abstract}
\section*{Abstract}

Polar amplification describes the predicted reduction in the latitudinal surface temperature gradient, which will have physical implications for mid-latitude dynamics.
The precise nature of these dynamical consequences remains unclear.
Here we explore aspects of polar amplification by way of 24 distinct idealised laboratory experiments.
The apparatus employed can independently prescribe laboratory analogues for the latitudinal temperature gradient ($\Delta{T}$, which controls the stratification $N$), background zonal flow speed ($U_{b}$), and strength of the background gradient in potential vorticity ($\beta$).
The ability to control these processes individually is beneficial as decoupling them from one another enables their influences can be examined separately.
Reducing the sidewall temperature difference substantially reduces small-scale and high frequency dynamics, but does not affect the large scale features of the flow, including the north-south amplitude of standing meanders.
Reducing the zonal flow speed does reduces the length-scales and amplitudes of the standing Rossby waves, while reducing the potential vorticity gradient has the opposite effect; these responses are well described by the canonical expression relating the standing Rossby wavelength to $\sqrt{U_{b}/\beta}$.
Variability is partitioned into components that are standing and transient; the response of this variability partitioning depends on all 3 experimental parameters, and a non-dimensional term is developed ($U_{b}\beta/N^{2}$) which captures the behaviour of the variability.
These findings suggest that the dynamical consequences of polar amplification is a tendency for mid-latitude weather to shift away from transient storms towards more persistent events, however the zonal wavelength and north-south extent of these persistent events will tend to decrease.

\end{abstract}

\begin{keywords}

differentially-heated rotating annulus; Rossby waves

\end{keywords}

\section{Significance Statement}

This study aims to improve our understanding of the dynamical consequences of polar amplification, which is a characteristic climate response to current and projected global warming scenarios.
Our approach is by way of laboratory experiments conducted in a large donut-shaped rotating tank with inner and outer sidewalls held at different temperatures, and in which we can impose flow-topography interactions and excite large-amplitude planetary waves.
By independently varying the tank rotation rate, flow speed, and sidewall temperature difference, we are able to test hypotheses regarding the planetary waves response to polar amplification.
Our findings suggest that polar amplification will lead to fewer short-lived storms but more persistent extreme weather events, and that the north-south amplitude of the planetary waves will decrease.


\section{Introduction}

Polar amplification -- the phenomenon whereby the Earth's surface at higher latitudes tend to warm more rapidly than lower latitudes under current and projected global warming scenarios -- is a robust response of Earth's climate to anthropogenic emissions of carbon dioxide \cite[e.g.,][]{bindoff_etal2013}.
Polar amplification was first observed in the Arctic \cite[e.g.,][]{serreze_francis2006, serreze_etal2009, screen_simmonds2010}, and occurs in the Antarctic but to a lesser extent, with the hemispheric asymmetry attributed to the different role of the polar oceans in the Northern and Southern Hemispheres \cite[e.g.,][]{marshall_etal2014, salzmann2017, stuecker_etal2018, smith_etal2019}.
Physically, polar amplification represents a reduction in the latitudinal temperature gradient, which subsequently reduces the baroclinicity and zonal wind speeds at mid-latitudes.
While there is scientific consensus that polar amplification is occurring now and will continue into the future, the dynamical consequences of reducing the mid-latitude baroclinicity and zonal wind speeds remains an open question.
Specifically, the effects that polar amplification will have on mid-latitude Rossby wave behaviours, such as their north-south amplitude, zonal wavelength, propagation speed, and general influence over mid-latitude weather, remain unclear.

Rossby waves play an important role in persistent extreme mid-latitude weather events \cite[e.g.,][]{hoskins_woollings2015, kornhuber_etal2017}.
Their dynamics are governed by several factors including the zonal wind speed $U$, baroclinicity or stratification arising from the latitudinal temperature difference $\Delta{T}$, and meridional gradient of potential vorticity $\beta$.
The fact that the phase velocity of Rossby waves is always oriented to the west means that eastward zonal flows can support arrested or near-stationary waves; in these case the wavelength of the arrested Rossby wave $\lambda$ is given by,
\begin{equation}
\lambda = 2\pi\sqrt{\frac{U}{\beta}}.
\label{eqn:lambda}
\end{equation}
Processes which influence these factors, will likely alter Rossby wave dynamics and, subsequently, mid-latitude weather \cite[e.g.,][]{screen_simmonds2013}.
Developing insight into the nature of these changes is important for understanding current observations and future projections of mid-latitude atmospheric dynamics.
For example, \cite{francis_vavrus2012} hypothesise that polar amplification will lead to a slower progression of Rossby waves and an increased north-south amplitude; these are effects which increase the probability of more prolonged extreme weather events such as drought, flooding, heat waves, and cold spells.
The \cite{francis_vavrus2012} hypothesis of an increased north-south amplitude of Rossby waves, and the subsequent implications for mid-latitude weather, has motivated considerable research into the subject, however without reaching scientific consensus \cite[e.g.,][]{hassanzadeh_etal2014, screen_simmonds2014, francis_vavrus2015, hoskins_woollings2015, barnes_screen2015, cattiaux_etal2016, blackport_screen2020}.

Part of the difficulty in testing the \cite{francis_vavrus2012} hypothesis with models stems from the coupled nature of the problem; the baroclinicity and zonal wind speed $U$ both depend on the latitudinal temperature difference $\Delta{T}$, which in turn depends on the baroclinicity and zonal wind speeds.
The meridional gradient of potential vorticity $\beta$, typically approximated by the latitudinal gradient of the Coriolis parameter $f$ (i.e., $\beta\approx{\partial{f}}/{\partial{y}}$), can change if the slope of the tropopause height were to change, which in turn is a response to the changing meridional temperature difference \cite[e.g.,][]{meng_etal2021}.
Furthermore, developing a model that is capable of supporting Rossby wave dynamics while remaining sufficiently baroclinic is challenging \cite[e.g.,][]{francis2017, vavrus2018, woollings_etal2018}; Rossby waves exist at the planetary scale, but are energised at the geostrophic scale, and decay at the turbulent viscosity scale via instabilities or wave breaking, thus requiring a wide range of active dynamical scales in the model to resolve crucial eddy-mean flow interactions \cite[e.g.,][]{lu_etal2015}.
In addition to this wide range of horizontal scales, the baroclinicity of the system requires the model to also have sufficient vertical resolution, where the definition of ``sufficient'' is not generally known {\it{a priori}}.
Even if such a {\it{numerical}} model were able to be developed, a reasonable exploration through dynamical parameter space to evaluate geophysical relevance would be computationally expensive.

In contrast, laboratory experiments with differentially-heated rotating annular tanks offer a powerful, convenient, and relatively inexpensive approach to investigating baroclinic Rossby wave dynamics \cite[e.g.,][]{read_etal2014, vincze_janosi2016}.
These simple annular configurations, referred to as ``Hide Tanks'' after \cite{hide1958}, are able to impose the fundamental physical conditions required to excite and maintain mid-latitude dynamics such as Rossby waves, zonal jets, baroclinic instability, and geostrophic/zonostrophic turbulence \cite[e.g.,][]{wordsworth_etal2008, smith_etal2014, rodda_etal2022, harlander_etal2023}.
One major scientific advantage that these Hide Tank experiments have over numerical simulations or global climate models is the ability to independently control the governing dynamical conditions of the system, allowing the sensitivities of specific processes to one another to be examined directly.
The recent renaissance of laboratory experiments is in part inspired by the ready access to modern technologies and analysis techniques providing new insights into processes that remain challenging and/or infeasible to explore via alternative approaches \cite[e.g.,][]{sutherland_etal2015}.

Here we present findings from a set of 24 distinct experiments using the Large Rotating Annulus (LRA) in the Climate \& Fluid Physics Laboratory at the Australian National University.
We are able to independently control the eastward zonal flow speed, strength of the background gradient of potential vorticity, and the sidewall temperature difference (i.e., the baroclinicity).
The experiments in focus here are a subset of a comprehensive suite of experiments that span a wide range of parameter space which covers $f$-plane and $\beta$-plane configurations, eastward and westward zonal flows, baroclinic and barotropic cases, and positive and negative latitudinal temperature gradients \citep{stewart_etal2025}.
The paper is laid out as follows: in \S2 we describe the apparatus, methodology, and analysis; in \S3 we present the experimental results; in \S4 we provide commentary on the geophysical implications of our findings in terms of the dynamical consequences of Polar Amplification; and conclude our study in \S5.

\section{Experiments}

\begin{figure}
\centering
\includegraphics[width=0.6\textwidth]{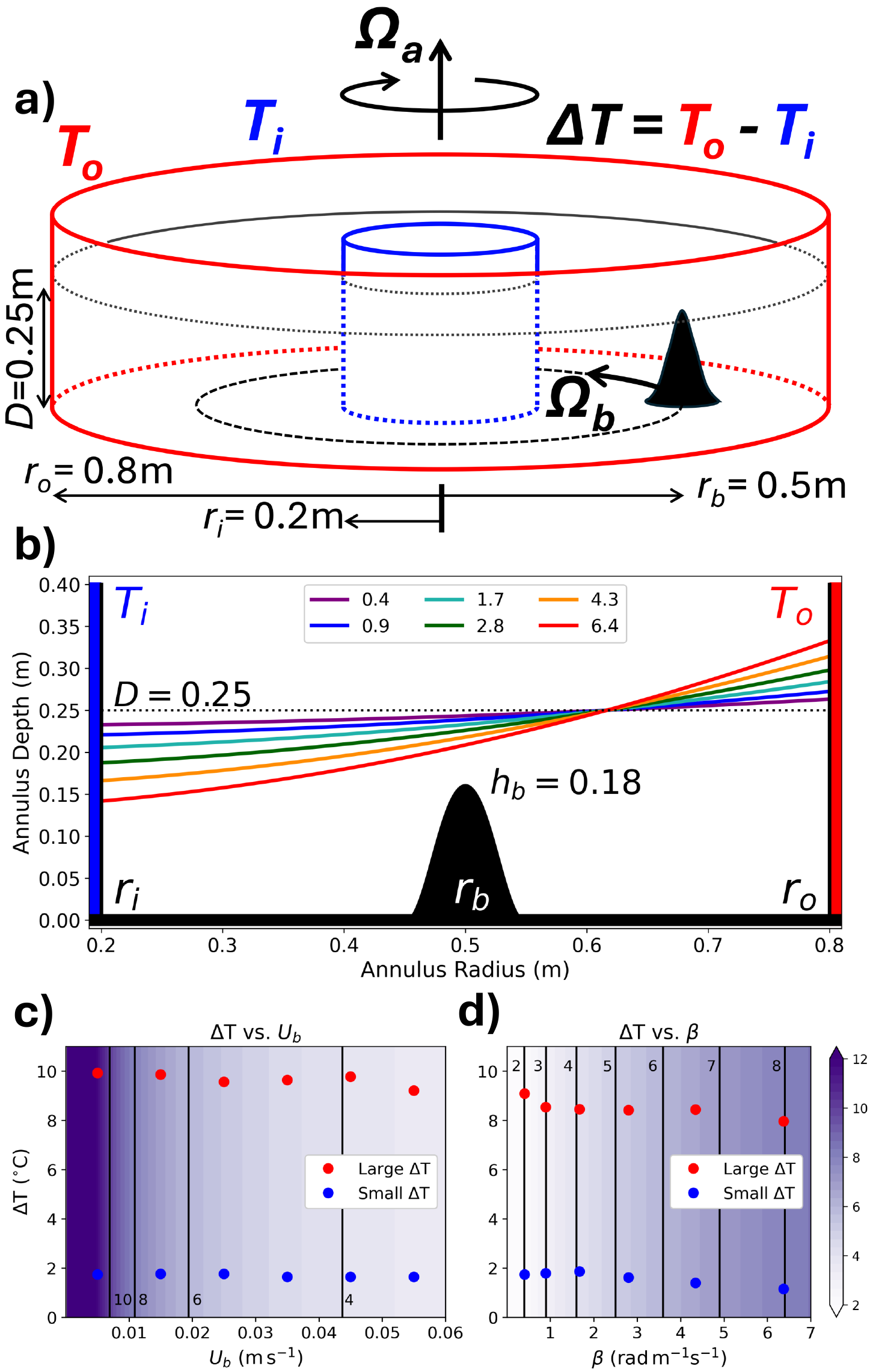}
\caption{Panel a) is a schematic of the Large Rotating Annulus (LRA). Our experiments vary the annulus rotation rate $\Omega_{a}$ (clockwise), the topography rotation rate $\Omega_{b}$ (anti-clockwise), and sidewall temperature difference $\Delta{T}$. Here, $\Omega_{a}$ varies the potential vorticity gradient $\beta$ according to Equation \ref{eqn:beta}, and $\Omega_{b}$ varies the zonal flow speed $U_{b}$ relative to the bump as $U_{b}=r_{b}\Omega_{b}$. Panel b) is a schematic of a radial transect across the LRA. The resting reference depth is represented by the horizontal dotted line at $D=$0.25\,m; the parabolic free surfaces of the different $\beta$ cases are shown by the solid coloured lines. A cross-section of topographic bump is included for scale. Panels c) and d) show the parameter spaces for experiments that vary the zonal flow speed $U_{b}$ and potential vorticity gradient $\beta$, respectively. In both cases, 2 sidewall temperature difference values are used: $\Delta{T}>8^{\circ}$C and  $\Delta{T}\approx2^{\circ}$C, referred to as large and small $\Delta{T}$, respectively. The background colours and black contours indicate the predicted zonal modenumber of the standing Rossby waves based on the respective $U_{b}$ and $\beta$ values.}
\label{fig:01_schematic}
\end{figure}

Here we briefly describe several key aspects of the Large Rotating Annulus (LRA) that are important for our analysis; comprehensive descriptions of the LRA are given in \cite{stewart_shakespeare2024} and \cite{stewart_etal2025}.
The LRA is a 1.6\,m diameter annular tank mounted atop a platform that rotates clockwise at a rate of $\Omega_{a}$ and filled with water to a resting reference depth of $D$=0.25\,m (Fig. \ref{fig:01_schematic}a).
The inner and outer sidewall temperatures of the LRA are independently controllable and maintained at temperatures $T_{i}$ and $T_{o}$, respectively; here $T_{o}>T_{i}$ such that $\Delta{T}=T_{o}-T_{i}>0$.
This sidewall temperature difference $\Delta{T}$ sets an upper bound on the stratification, which can be expressed as the Brunt-V{\"a}is{\"a}l{\"a} buoyancy frequency $N$,
\begin{equation}
N = \sqrt{g\alpha\frac{\Delta{T}}{D}},
\label{eqn:buoyancy}
\end{equation}
where $g$ is gravity and $\alpha$ is the thermal expansion coefficient.
A three-dimensional gaussian isolated topographic bump (height $h_{b}$=0.18\,m, half-width $w_{b}$=0.1\,m) is located at the mid-radius $r_{b}$=0.5\,m of the LRA and is able to be differentially rotated around the annulus at a rate $\Omega_{b}$; here $\Omega_{b}$ is anticlockwise.
In the reference frame of the topographic bump, the water moves clockwise or eastward past the bump at a speed of $U_{b}=r_{b}\Omega_{b}=0.5\Omega_{b}$.

When the LRA is rotating, the water experiences a radially-outward centrifugal force that becomes balanced by a radially-inward hydrostatic pressure force.
This force balance results in the free surface depth $d$ becoming parabolic with radius and dependent on $\Omega_{a}$ (Fig. \ref{fig:01_schematic}b).
This radial dependence of depth is dynamically equivalent to a radial gradient of potential vorticity $\beta$ which can be expressed as,
\begin{equation}
\beta = \frac{\Omega_{a}^{3}r_{b}}{gD}.
\label{eqn:beta}
\end{equation}
Thus, by changing the LRA rotation rate $\Omega_{a}$, we are able to vary the strength of the background gradient of potential vorticity $\beta$.

The primary diagnostic tool employed here is a wide-angle (42$^{\circ}\times$32$^{\circ}$) FLIR E75 thermal camera centrally-located above the LRA on a frame that co-rotates with the table.
The spatial resolution of the camera is at least $5\times5$\,mm$^2$, and the field of view is able to cover approximately 95\% of the water surface.
All of the analysis presented here is performed in the frame of reference of the topographic bump, such that the thermal images from the FLIR camera are rotated about their centre by an angle that depends on the bump rotation rate $\Omega_{b}$.
This process transforms the thermal images into a frame of reference where the bump is stationary and the water flows zonally eastward past the bump.
All cases use 2 hours of thermal camera footage recorded at 1\,Hz, so 7200 images.
An optical camera is also mounted above the LRA and co-rotates with the table; this is used for visualising the flow with passive dye tracer.

The 3 independent parameters we vary here are the eastward zonal flow speed $U_{b}$, the background gradient of potential vorticity $\beta$, and the sidewall temperature difference $\Delta{T}$.
We present cases that impose 6 different eastward flow speeds $U_{b}$ spanning 5--55\,mm\,s$^{-1}$ at 10\,mm\,s$^{-1}$ intervals (Fig. \ref{fig:01_schematic}c), and 6 different values of $\beta$ (Fig. \ref{fig:01_schematic}d).
From Equation \ref{eqn:lambda}, these values of $U_{b}$ and $\beta$ are expected to support standing Rossby waves with modenumbers from 2 up to approximately 12 (contours on Fig. \ref{fig:01_schematic}c,d).
Each $U_{b}$ and $\beta$ experiment is conducted for 2 different sidewall temperature differences $\Delta{T}$; one small ($\Delta{T}<2^{\circ}$C) and one large ($\Delta{T}>8^{\circ}$C) temperature difference.

\section{Results}

We first present qualitative descriptions of the flows and the observed effects of the temperature difference $\Delta{T}$, zonal flow speed $U_{b}$, and the background gradient of potential vorticity $\beta$.
These qualitative descriptions are based on visualisations of the passive dye release experiments, and the thermal structures observed with the FLIR camera.
We then use the thermal camera data for subsequent quantitative analysis; we compare the influence of the 3 experimental parameters ($\Delta{T}$, $U_{b}$, $\beta$) on the waviness of the flow (as given by the radial variability of the thermal structure), the nature of the variability (standing or transient), and its wavelength and frequency composition.

\subsection{Qualitative Description}

\subsubsection*{Passive Dye Visualisations}

\begin{figure}
\centering
\includegraphics[width=1.0\textwidth]{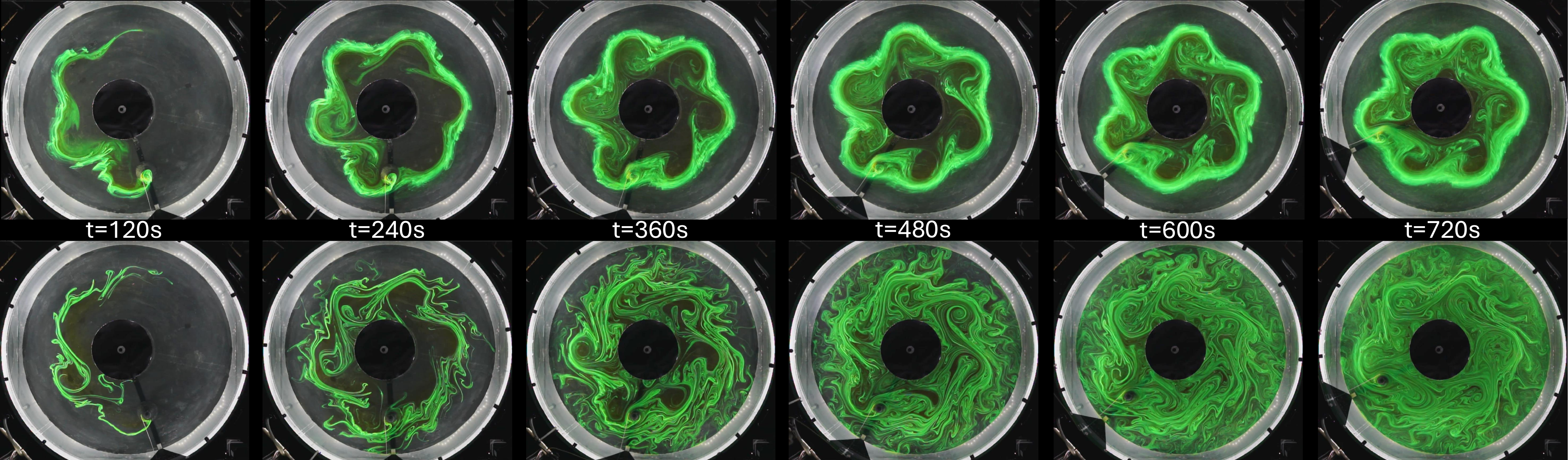}
\caption{A sequence of images taken at regular time intervals after passive tracer dye is released into the equilibrated flows of experiments with $U_{b}=$25\,mm\,s$^{-1}$ and $\beta=2.8$\,rad\,m$^{-1}$\,s$^{-1}$, for small and large sidewall temperature differences (top and bottom rows, respectively). Note that the large scale structures of the different cases are remarkable similar; however, the larger $\Delta{T}$ case exhibits substantially more small scale features and distribution of the passive tracer. These images are available in video form on the {\it{FluidsIn4K}} youtube channel at {\it{www.youtube.com/watch?v=CatIj6DU6ss}}}
\label{fig:02_dye_release}
\end{figure}

Passive dye tracer was introduced into equilibrated experiments by way of a thin tube attached to the bump.
These dye tracer visualisation runs provide an opportunity to directly view the flow structure and characteristics, and compare these across different experiments.
Figure \ref{fig:02_dye_release} shows a pair of photo sequences taken at 120\,second intervals in the frame of reference of the annulus.
These photo sequences are for experiments that have identical zonal flow speed $U_{b}=25$\,mm\,s$^{-1}$ and potential vorticity gradient $\beta=2.8$\,rad\,m$^{-1}$\,s$^{-1}$; for this value of $U_{b}$ the bump completes an anticlockwise lap around the annulus every 125.6\,seconds.
The case shown in the upper row has the smaller sidewall temperature difference of $\Delta{T}\approx{2}^{\circ}$C, while the lower row has the larger $\Delta{T}\approx{8}^{\circ}$C; this is the only difference between these two cases.

Evident in both cases are zonal jet-like structures with distinct zonal modenumber 6 meanders.
This zonal jet exhibits notable consistency between the two experiments; the flow radially inwards the jet occurs as 6 distinct cyclonic (clockwise) features, and the phase of the jet meanders appears to be locked to the bump.
The similarity of the jets and their modenumber 6 structure is made more remarkable because of the obvious differences in the smaller scale features of the flows; the case with the larger temperature difference exhibits substantially more small-scale eddies and filaments, and a broader distribution of passive dye tracer throughout the annulus.
Indeed, in the case of the smaller temperature difference, the radial extent of the passive dye tracer appears to be bound by the jet, and concentrated within the jet itself.
That is, for the smaller temperature difference case, the jet appears to prevent tracer reaching the outer regions of the annulus; there is no such dynamical boundary or constraint for the larger temperature difference case.

\subsubsection*{Zonal Flow Speed $U_{b}$}

\begin{figure}
\centering
\includegraphics[width=1.0\textwidth]{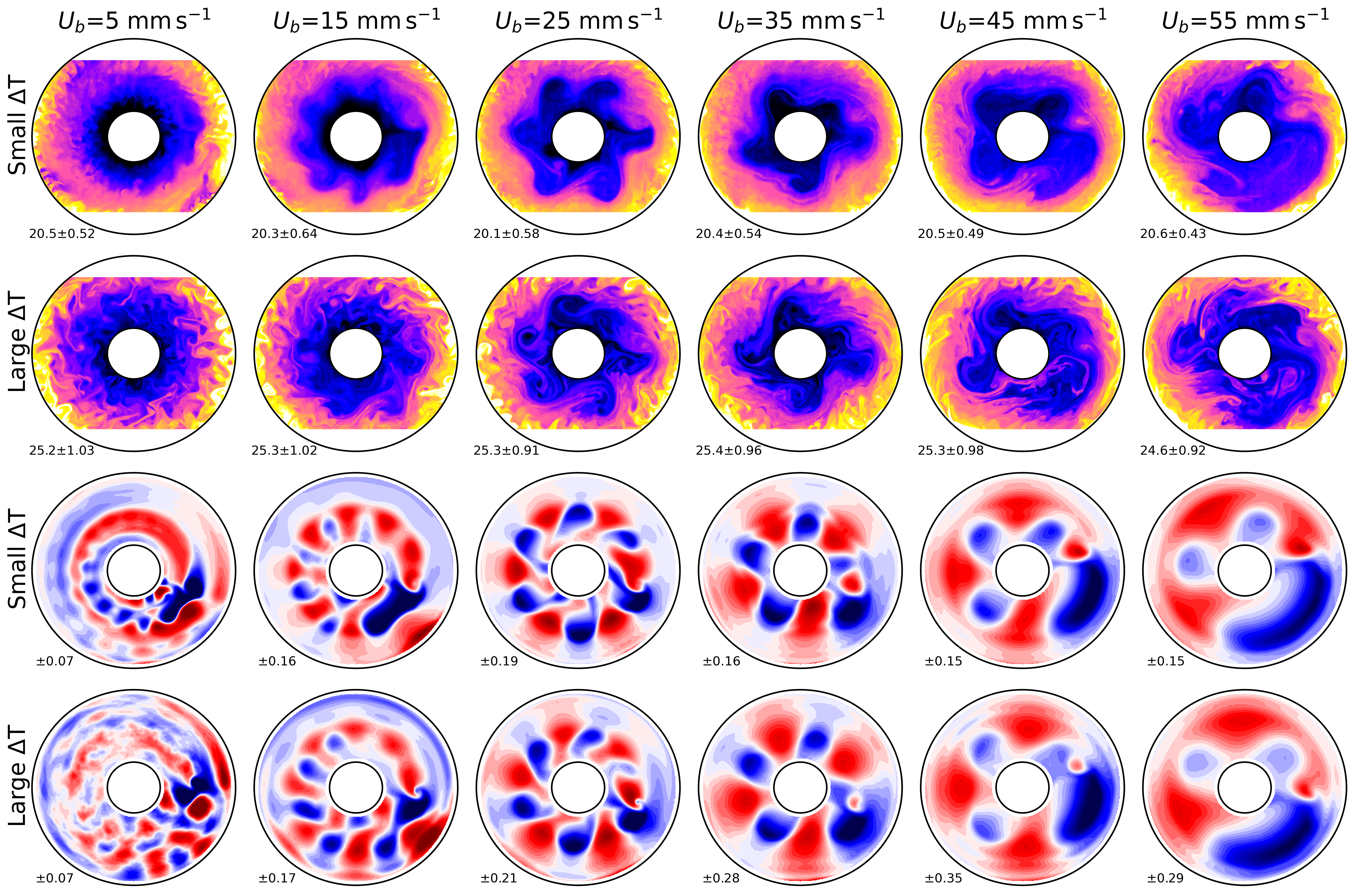}
\caption{The upper 2 rows show snapshots of surface temperatures for experiments that vary the zonal flow speed $U_{b}$ with small and large sidewall temperature differences $\Delta{T}$ (top and second row, respectively). The colourmap in each panel is centered about the snapshot mean of the annulus temperature and spans a range given by the snapshot standard deviation of the annulus temperature; these values are indicated in the lower left of each panel. The lower 2 rows show the time averaged surface temperature anomalies relative to the location of the bump for the same set of experiments as the upper 2 rows. The colourmaps are centered about zero and span a range given by the annulus standard deviation, which is indicated in the lower left of each panel.}
\label{fig:03_SST_snaps_means_ub}
\end{figure}

The flow structures made visible by the passive dye tracer are consistent with the surface temperature distributions obtained by the thermal camera, and these thermal fields have the advantage of allowing further quantitative comparisons of the flows.
The upper two rows of Figure \ref{fig:03_SST_snaps_means_ub} shows snapshots of surface temperatures for the 6 values of $U_{b}$ and the 2 sidewall temperature differences $\Delta{T}$.
The colourmaps in each case are centered about the annulus mean temperature of the snapshot and span a temperature range given by the standard deviation of temperature; these values are indicated in the lower left of each panel.
The reasoning for this varying colourmap is that we aim to highlight the nature of the thermal structures in each case, rather than the absolute temperature fields. 
All cases have the characteristic thermal structures consistent with geostrophic turbulence, and exhibit a meandering thermal front at mid-radius.
The large scale meanders of the thermal front appear to depend on the zonal flow speed, with the wavelength of the meanders increasing with $U_{b}$.
The structure of the large scale meanders appear insensitive to the sidewall temperature difference.
The large $\Delta{T}$ cases exhibit substantially more smaller scale features, including eddies and filaments, especially closer to the outer sidewall.

The time average of the 2 hours of surface temperatures are used to obtain a timeseries of surface temperature anomalies; these are then rotated into the frame of reference of the topographic bump, and averaged in time.
The bottom two rows of Figure \ref{fig:03_SST_snaps_means_ub} shows these standing temperature anomalies; this analysis highlights thermal features that are stationary relative to the topographic bump.
Again, the colourmaps are centered about the annulus mean (which is by definition zero), and spans a temperature range given by the standard deviations in each case; these are indicated to the lower left of each panel.
All cases exhibit regular alternating positive and negative anomalies emanating from topographic bump.
The anomalies appear of similar shape to the large scale meanders of the respective snapshots; the length-scale of these anomalies increases with $U_{b}$.
The phase of the anomalies appears to be set by the topographic bump; in the vicinity of the bump, the anomalies are warm upstream (anticlockwise) and cool downstream (clockwise).
The radial extent of these standing anomalies appears consistent with the zonal wavelength, and insensitive to the sidewall temperature difference.

\begin{figure}
\centering
\includegraphics[width=1.0\textwidth]{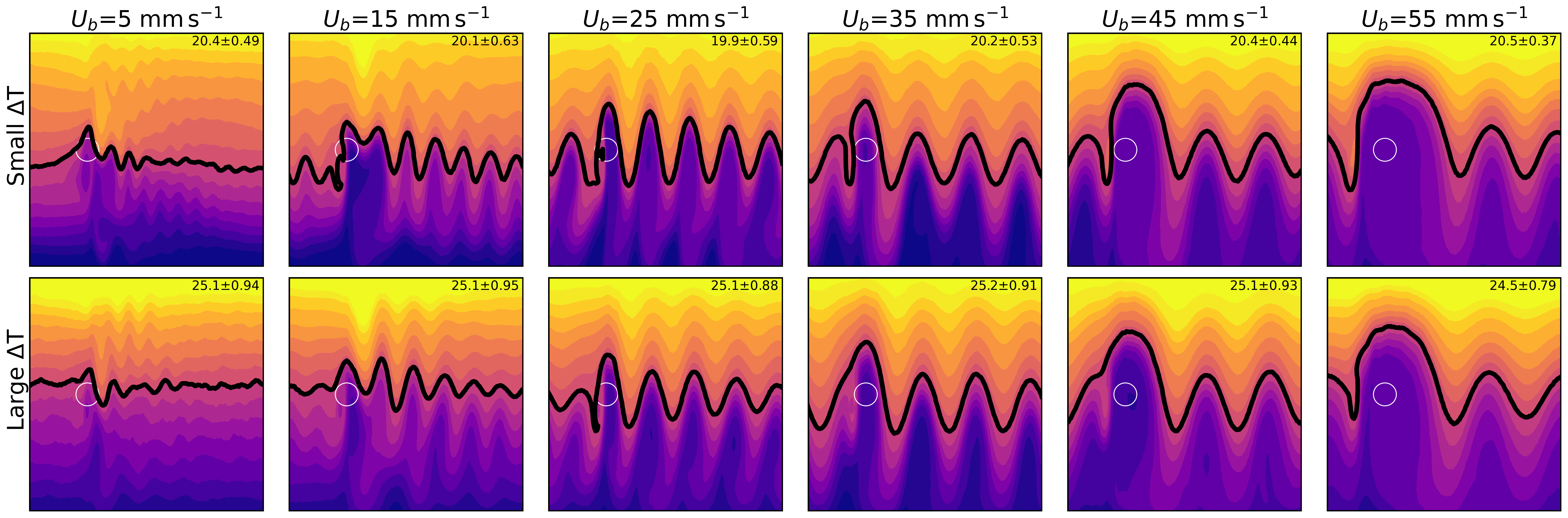}
\caption{Time average surface temperatures of the experiments that vary $U_{b}$ for small and large $\Delta{T}$ (top and bottom rows, respectively); here the temperature fields have been transformed into radial--azimuthal coordinates. The annulus mean temperature of each case is contoured in black, and the approximate location of the bump is indicated by the white circle. The colourmap is centered about the annulus mean temperature and spans a range given by the standard deviation; these values are included in the upper right of each panel.}
\label{fig:04_zonal_means_ub}
\end{figure}

The meandering structure of the temperature fields is more apparent when transformed into radial--azimuthal coordinates.
Figure \ref{fig:04_zonal_means_ub} shows the time average of surface temperatures relative to the bump in radial--azimuthal coordinates.
In each case, the annulus mean temperature is shown by the black contour, and the approximate location of the bump by the white circles; again, the colourmaps are centered about the mean temperature $\overline{T}$ and span a range given by the standard deviation, which are listed in the upper right of each panel.
The standing meanders are obvious; these are of largest radial extent immediately at the bump, and tend to decay downstream (east).
The wavelength of the meanders increases with zonal flow speed $U_{b}$.
The large scale structure of the meanders appears insensitive to the sidewall temperature difference $\Delta{T}$.

\subsubsection*{Potential Vorticity Gradient $\beta$}

\begin{figure}
\centering
\includegraphics[width=1.0\textwidth]{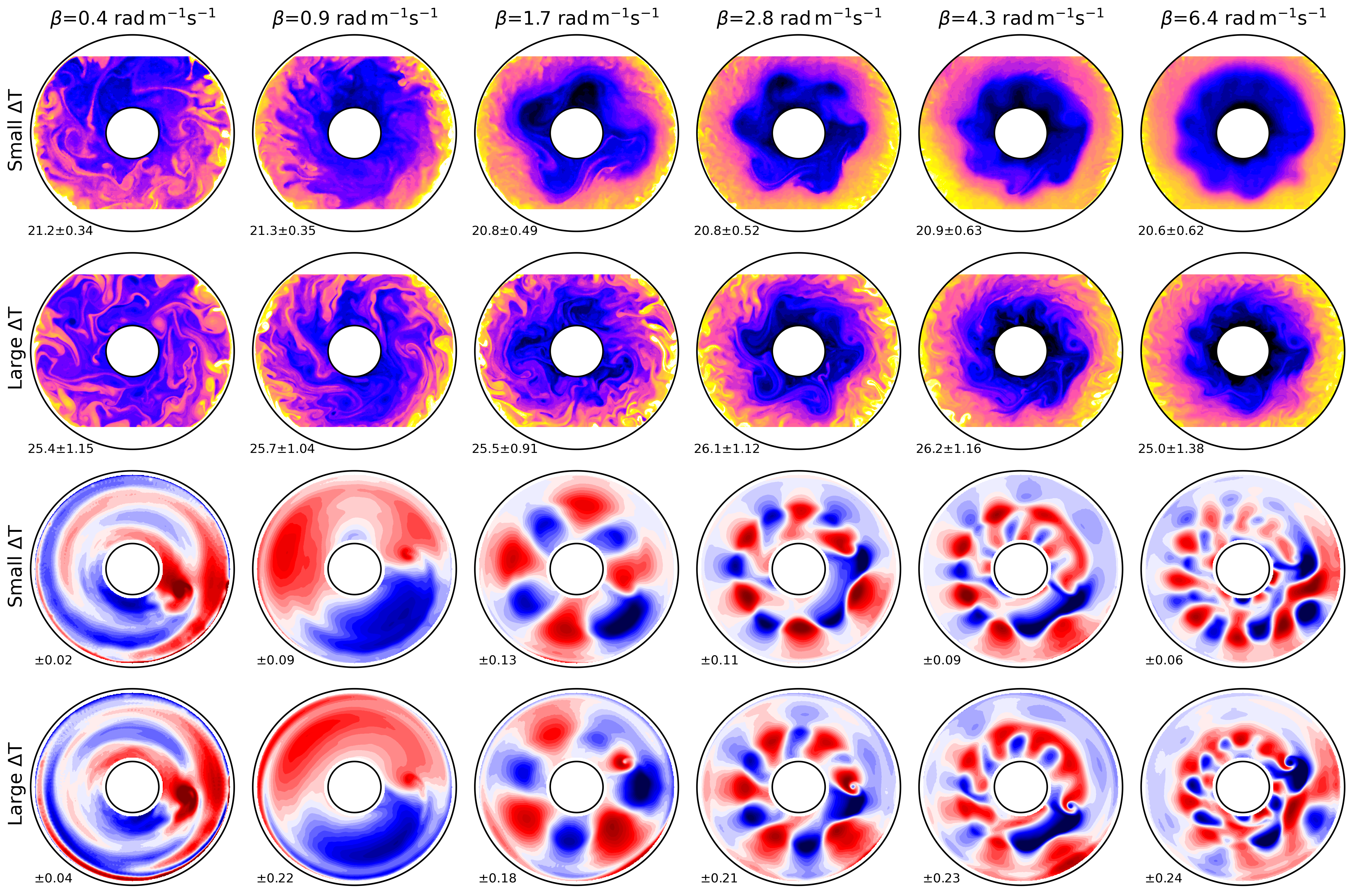}
\caption{The effect of varying the potential vorticity gradient $\beta$ on surface temperature snapshots (upper 2 rows) and standing temperature anomalies (lower 2 rows); these panels have the same layout and colourmap convention as per Figure \ref{fig:03_SST_snaps_means_ub}.}
\label{fig:05_SST_snaps_means_beta}
\end{figure}

The thermal structures of the flow are sensitive to the strength of the potential vorticity gradient $\beta$.
Figure \ref{fig:05_SST_snaps_means_beta} shows snapshots of surface temperatures (top two rows) and their corresponding standing temperature anomalies (bottom two rows) for the 6 values of $\beta$ and 2 sidewall temperature differences $\Delta{T}$; the colourmaps follow the same convention as for Figure \ref{fig:03_SST_snaps_means_ub}.
The cases with $\beta<1$\,rad\,m$^{-1}$\,s$^{-1}$ exhibit circulations rich with filaments and eddies, but without large scale zonal structures.
As $\beta$ increases beyond $\beta=1.7$\,rad\,m$^{-1}$\,s$^{-1}$, the temperature snapshots begin to feature the characteristic meandering thermal front evident in the $U_{b}$ experiments (Fig. \ref{fig:03_SST_snaps_means_ub}).
The zonal wavelength of the meanders reduces with increasing $\beta$; notably, the radial extent of the meanders also reduces with increasing $\beta$.
The larger $\Delta{T}$ cases all exhibit relatively more small scale dynamics, but are generally consistent with their corresponding smaller $\Delta{T}$ in terms of the large scale structures.
The standing temperature anomaly fields further highlight the remarkable similarity between the different $\Delta{T}$ cases; the structure of the thermal anomaly fields are insensitive to the sidewall temperature difference $\Delta{T}$.
The zonal wavelength of the standing temperature anomalies decreases with increasing $\beta$; so too the radial extent of these standing anomalies.

\begin{figure}
\centering
\includegraphics[width=1.0\textwidth]{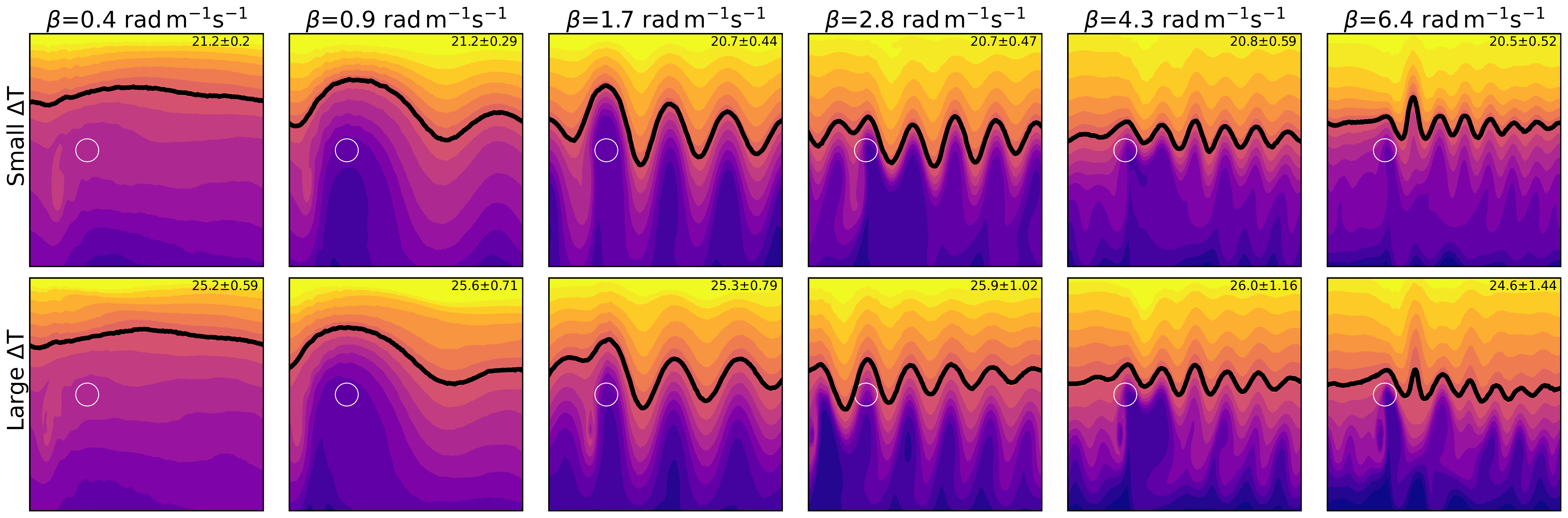}
\caption{The effect of varying the potential vorticity gradient $\beta$ on the time average surface temperatures; these panels have the same layout and colourmap convention as per Figure \ref{fig:04_zonal_means_ub}.}
\label{fig:06_zonal_means_beta}
\end{figure}

The characteristic nature of the standing meanders is again more apparent when transformed into radial--azimuthal coordinates (Fig. \ref{fig:06_zonal_means_beta}).
The zonal wavelength and radial extent of the meanders decrease as $\beta$ increases.
The radial extent of the standing meanders is largest in the immediate vicinity of the bump, and appears to decay downstream (east) past the bump.
The standing thermal structures appear insensitive to the sidewall temperature difference $\Delta{T}$.

\subsection{Quantitative Analysis}

\begin{figure}
\centering
\includegraphics[width=1.0\textwidth]{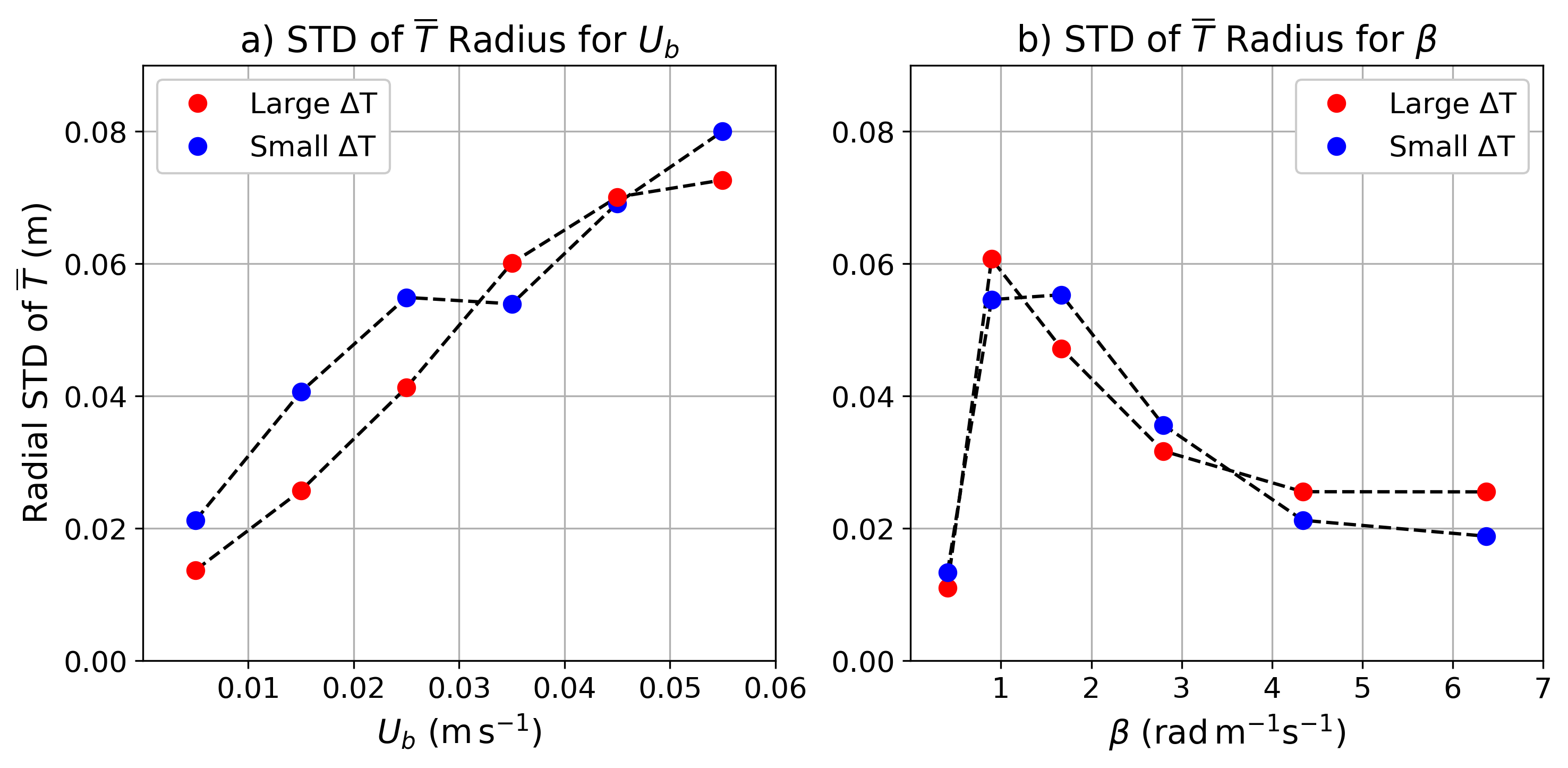}
\caption{The standard deviation of the radial location of the annulus mean isotherm for each experiment (i.e., the black contours shown in Figures \ref{fig:04_zonal_means_ub} and \ref{fig:06_zonal_means_beta}); this metric provides a simple measure of the relative waviness of each experiment. Panels a) and b) show the effects the zonal flow speed $U_{b}$ and potential vorticity gradient $\beta$, respectively.}
\label{fig:07_jet_std}
\end{figure}

We are interested in understanding the effect of the zonal flow speed $U_{b}$, potential vorticity gradient $\beta$, and sidewall temperature difference $\Delta{T}$ on the radial extent of the standing meanders.
For this, we require a metric that quantifies the radial waviness of the meanders, akin to a sinuosity index \cite[e.g.,][]{cattiaux_etal2016}.
The annulus mean isotherm contoured in Figures \ref{fig:04_zonal_means_ub} and \ref{fig:06_zonal_means_beta} appear to be simple indicator of the waviness of the mean temperature field; this isotherm also tends to reflect the approximate location of the jet in each experiment.
The standard deviation of the radial position of the annulus mean isotherm provides a measure of the radial waviness of the system; a small (large) standard deviation implies small (large) radial amplitude to meanders.
Figure \ref{fig:07_jet_std}a,b shows the standard deviations of the radial position of the annulus mean isotherm for the experiments with $U_{b}$ and $\beta$, respectively.
These measures of radial waviness increase with increasing zonal flow speed $U_{b}$, indicating that larger flow speeds have larger radial amplitude meanders; this is consistent with the qualitative structures evident in Figure \ref{fig:04_zonal_means_ub}.
The radial waviness generally decreases with increasing potential vorticity $\beta$; the exceptions here are for the weakest values of $\beta<1$, which do not appear to have a coherent jet or substantial standing thermal structures.
Again, the trends indicated by this radial waviness metric are consistent with the quantitative structures in Figure \ref{fig:06_zonal_means_beta}.
Notably, in both experiment sets, there is no obvious difference in these radial waviness metrics between cases with large and small sidewall temperature differences $\Delta{T}$.

\begin{figure}
\centering
\includegraphics[width=1.0\textwidth]{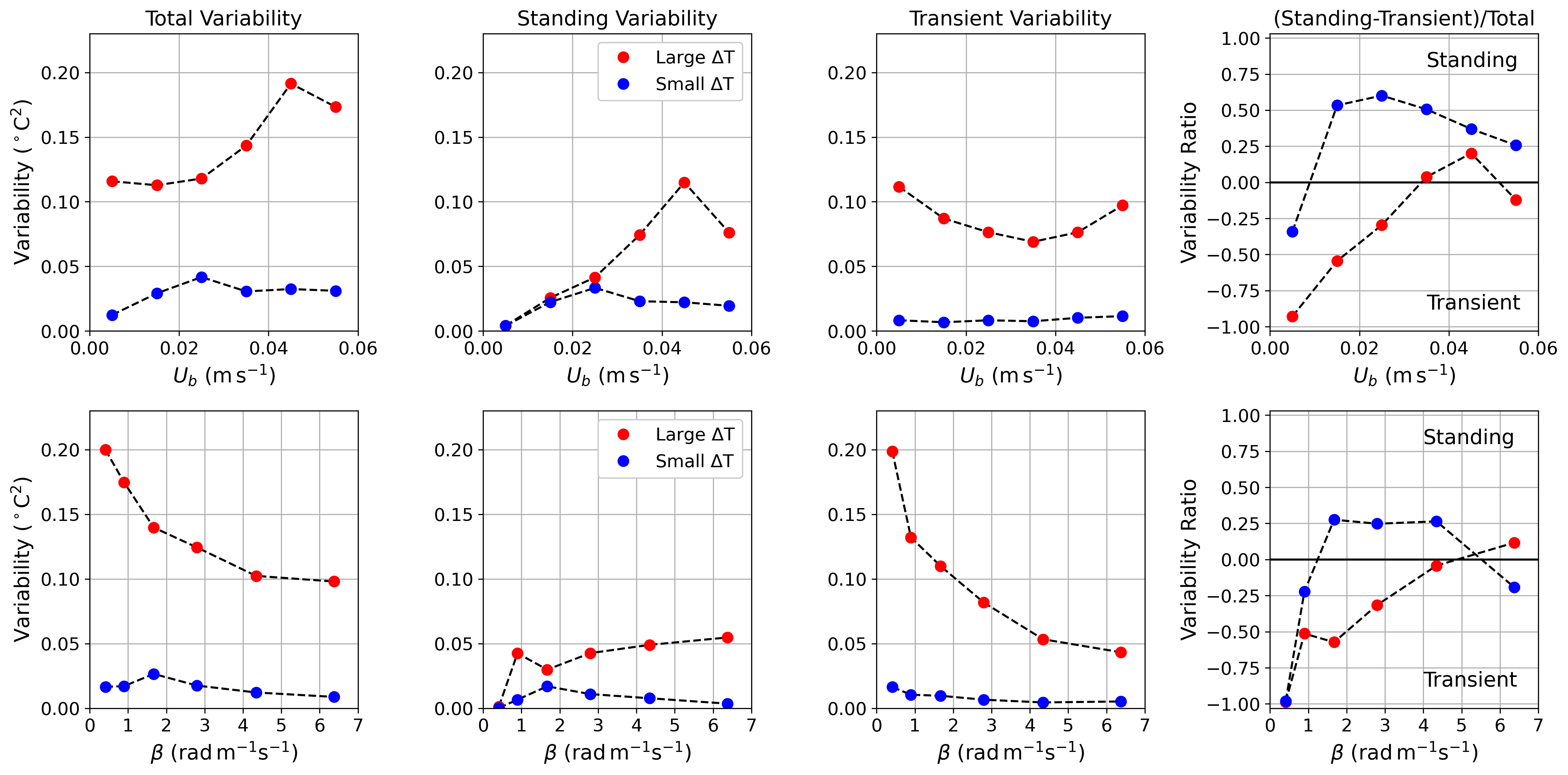}
\caption{Columns 1--3 show the total, standing and transient variabilities, respectively, for experiments that vary the zonal flow speed $U_{b}$ and $\beta$ (top and bottom rows, respectively). The rightmost column shows the variability ratio, which is defined as the ratio of the difference between the standing and transient variabilities to the total variability.}
\label{fig:08_variabilities}
\end{figure}

A Reynolds decomposition of the surface temperature anomalies can partition the thermal variability into a time-invariant component which is stationary relative to the bump (e.g., the time averaged temperature anomalies shown in the lower two rows of Figs. \ref{fig:03_SST_snaps_means_ub}, \ref{fig:05_SST_snaps_means_beta}), and time-dependent component.
These components are referred to here as the standing and transient variabilities, respectively, and their sum returns the total thermal variability of the system; specific details of this variability decomposition are provided in \cite{stewart_etal2025}.
The annulus averages of these variability components are shown in Figure \ref{fig:08_variabilities} for the experiments that vary $U_{b}$ (top row) and $\beta$ (bottom row).
The experiments with large $\Delta{T}$ exhibit more total thermal variability than their respective small $\Delta{T}$ cases; this makes good physical sense because the large $\Delta{T}$ cases have a wider distribution of temperatures, as indicated by the relative sizes of the standard deviations of temperature listed in Figures \ref{fig:03_SST_snaps_means_ub}--\ref{fig:06_zonal_means_beta}.
There is generally an increase in the total variability with increasing $U_{b}$, although the total thermal variability appears to saturate in experiments with the largest $U_{b}$ values; we hypothesise that this saturation is due to more mixing occurring in these cases, which leads to a reduction in range of temperature distributions, as indicated by the decreasing standard deviations listed for these cases in Figures \ref{fig:03_SST_snaps_means_ub}--\ref{fig:04_zonal_means_ub}.
In contrast to $U_{b}$, the total thermal variability decreases with increasing $\beta$; this is representative of the dynamical transition from geostrophic to zonostrophic turbulence as $\beta$ increases.

The standing variability for generally increases with $U_{b}$ and $\beta$, especially for the large $\Delta{T}$ cases; for $U_{b}\leq$25\,mm\,s$^{-1}$, the amount of standing variability appears insensitive to the sidewall temperature difference $\Delta{T}$.
One distinction to note about the effect of sidewall temperature difference is that for the small $\Delta{T}$ experiments, the amount of standing variability appears to saturate or decrease for larger $U_{b}$ and $\beta$ values; this is not the case for the large $\Delta{T}$ experiments.
The amount of transient variability exhibits a relatively more complicated dependence on $U_{b}$; for the large $\Delta{T}$ experiments, the transient variability initially decreases with increasing $U_{b}$ until reaching a minimum for $U_{b}=$35\,mm\,s$^{-1}$, and then increasing beyond that.
This minimum in transient variability for $U_{b}$ is not obvious in the small $\Delta{T}$ experiments.
In terms of $\beta$, there is a monotonic decrease in transient variability for increasing values of $\beta$; again, this is consistent with the dynamical transition from geostrophic to zonostrophic turbulence.

The rightmost column of Figure \ref{fig:08_variabilities} shows the ratio of the standing minus transient variabilities to the sum of the standing and transient variabilities, which is the total variability.
By definition, this ratio lies between -1 and 1; a ratio of -1 indicates the total variability is entirely composed of transient variability, 1 means the total variability is entirely standing variability, and 0 indicates equipartitioned variability.
The sidewall temperature difference $\Delta{T}$ has a distinct influence over the effect of the zonal flow speed $U_{b}$ on the variability ratio.
For large $\Delta{T}$, the variability ratio is predominantly transient for small $U_{b}$, increases with increasing $U_{b}$, and becomes approximately equipartitioned for $U_{b}>$35\,mm\,s$^{-1}$.
For small $\Delta{T}$, however, the variability ratio is predominantly standing for cases with $U_{b}>$5\,mm\,s$^{-1}$, and tends to decrease for increasing $U_{b}$; the only case with greater transient variability is that of $U_{b}=$5\,mm\,s$^{-1}$.
To reiterate, for decreasing $U_{b}$, the variability of large $\Delta{T}$ cases tend to become more transient, while for small $\Delta{T}$ case the variability tends to become more standing.

The effect of increasing $\beta$ on the variability ratio is comparatively simpler; the ratio increases with $\beta$ for both small and large $\Delta{T}$ from completely transient for $\beta=$0.4\,rad\,m$^{-1}$\,s$^{-1}$.
The small $\Delta{T}$ cases exhibit slightly more standing than transient variability for intermediate values of $\beta$, but remains approximately equipartitioned.
The large $\Delta{T}$ cases have relatively more transient variability than their respective small $\Delta{T}$ cases, and become weakly predominantly standing for the largest value of $\beta$; this is the only value of $\beta$ for which the large $\Delta{T}$ experiment is more standing than the respective small $\Delta{T}$ experiment.

\begin{figure}
\centering
\includegraphics[width=1.0\textwidth]{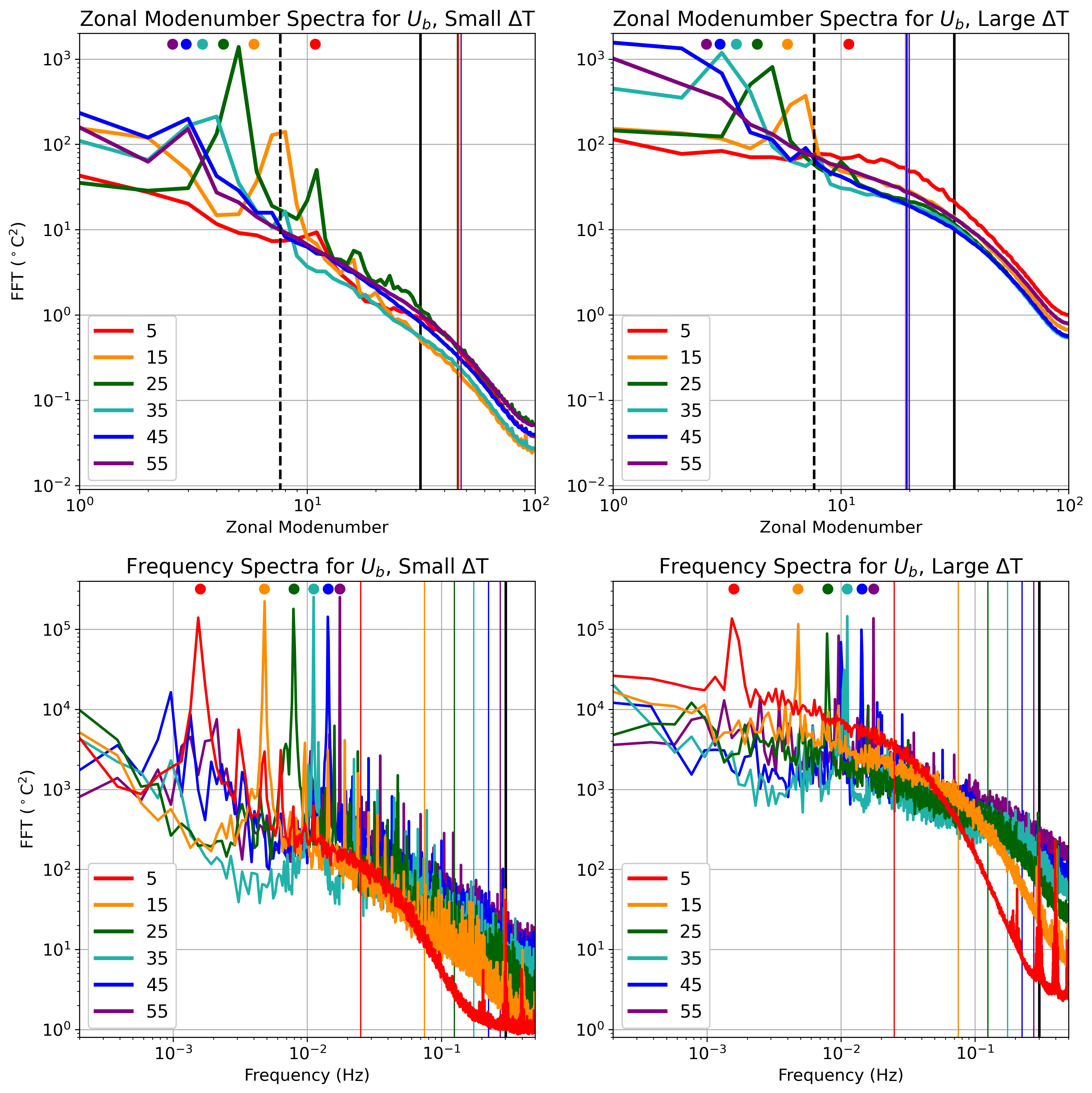}
\caption{Modenumber and frequency spectra (top and bottom rows, respectively) of the experiments that vary the zonal flow speed $U_{b}$ for small and large sidewall temperature differences (left and right columns, respectively). For the modenumber panels (top row): the coloured dots indicate the modenumber of the predicted standing Rossby waves for each case; the dashed and solid black lines represent the modenumbers of the barotropic Rossby deformation radius and the topography, respectively; and the thin coloured lines represent the baroclinic Rossby deformation radii for each case. For the frequency panels (bottom row): the coloured dots represent the return frequency of the topography; the solid black line represents the inertial frequency of the annulus; and the thin coloured lines represent the intrinsic frequency of the topography in each case.}
\label{fig:09_FFTs_ub}
\end{figure}

The variability of the systems can be further understood by comparing their wavelength and frequency spectra.
Figure \ref{fig:09_FFTs_ub} shows the zonal modenumber and frequency spectra (top and bottom rows, respectively) of the small and large $\Delta{T}$ cases (left and right columns, respectively) for the experiments that vary zonal flow speed $U_{b}$.
The modenumber of the topography, which represents the forcing length-scale of the system, is indicated by the solid black line; note that the baroclinic deformation radius modenumber is larger than the topography for the small $\Delta{T}$ cases, and smaller than the topography for the large $\Delta{T}$ cases.
The variability at a given modenumber is relatively insensitive to the zonal flow speed, with the exception of the distinct peaks in energy near the predicted standing Rossby wave modenumbers (coloured dots above the curves).
The large $\Delta{T}$ cases have relative more energy than their small $\Delta{T}$ case for almost all modenumbers; again, the exception is in the vicinity of the predicted standing Rossby wave modenumber, where the energies of the different $\Delta{T}$ cases are of similar magnitude.
The large $\Delta{T}$ cases exhibit a relatively flatter spectra through their baroclinic deformation radius modenumber and to the modenumber of topography; this is consistent with the observation of relatively more smaller scale structures in the dye tracer and thermal camera visualisations.
Note that all cases except $U_{b}=$5\,mm\,s$^{-1}$ have predicted standing Rossby wave modenumbers that are smaller than the modenumber of the barotropic Rossby deformation radius.

The frequency spectra exhibit relatively more sensitivity to the zonal flow speed $U_{b}$ than the modenumber spectra; increasing $U_{b}$ has an obvious tendency to increase the energy at intermediate frequencies.
The frequency spectra all have substantial spikes at their respective topography return frequencies (i.e., the frequency that represents the time taken for the topographic bump to complete a lap around the annulus: $U_{b}/2\pi{r_{b}}$), indicated by the coloured dots.
The large $\Delta{T}$ cases generally exhibit a larger amount of energy for a given frequency, with the exception of the topography return frequencies, but the general increase is greater for higher frequencies relative to lower frequencies.
The spectra drop-off to high frequency, which is steepest for small values of $U_{b}$, steepens for frequencies higher than the topographic intrinsic or forcing frequency (i.e., the frequency that represents the time taken for flow to travel past the topographic bump; $U_{b}/w_{b}$).
Overall, the effect of increasing zonal flow speed $U_{b}$ tends to leave the wavelengths of the variability unchanged, except for those associated with standing Rossby waves, and increase the amount of energy at higher frequencies.
The larger sidewall temperature difference $\Delta{T}$ tends to have more energy overall, especially at higher modenumbers and frequencies.

\begin{figure}
\centering
\includegraphics[width=1.0\textwidth]{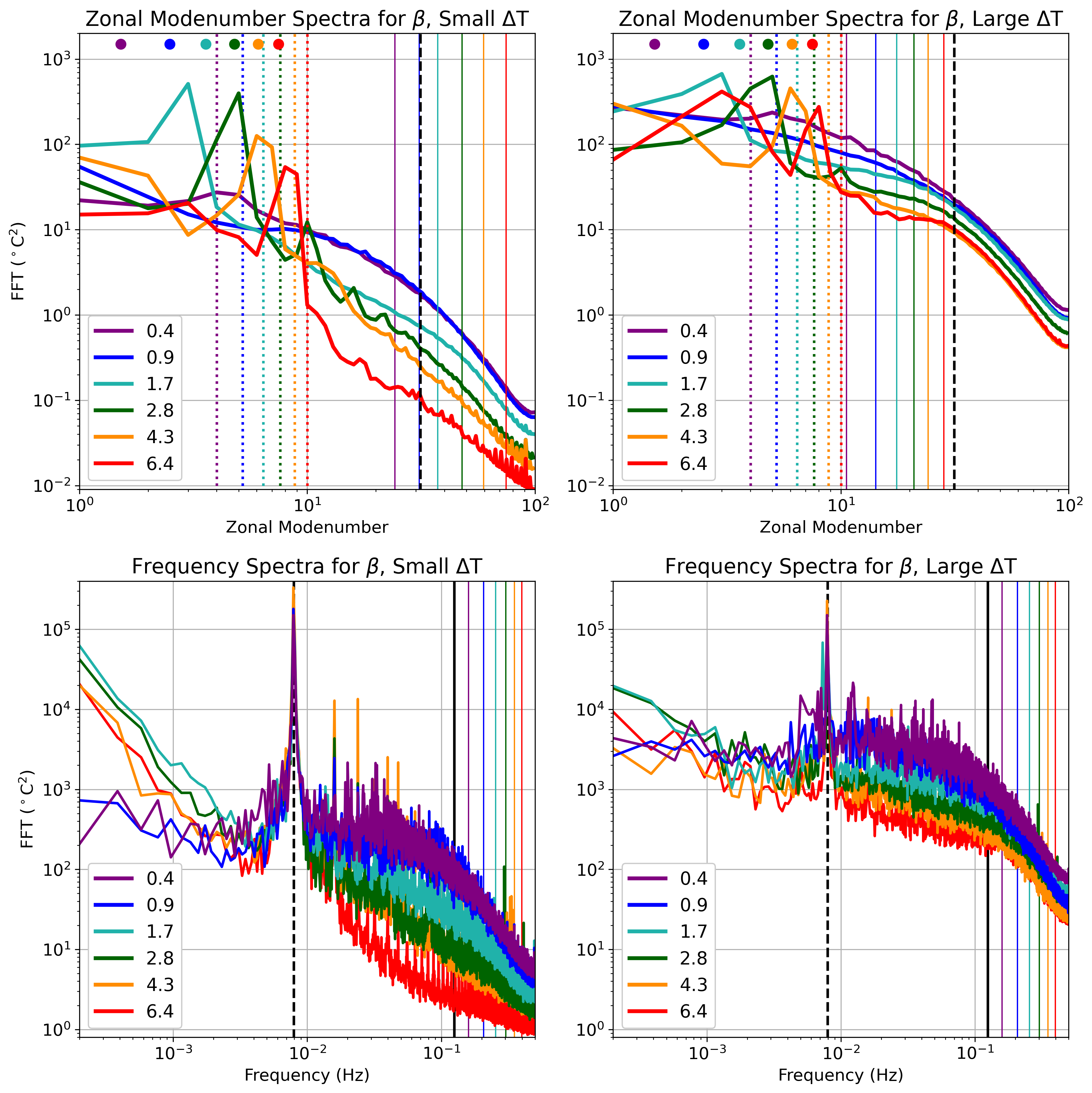}
\caption{Modenumber and frequency spectra (top and bottom rows, respectively) of the experiments that vary the potential vorticity gradient $\beta$ for small and large sidewall temperature differences (left and right columns, respectively). For the modenumber panels (top row): the coloured dots indicate the modenumber of the predicted standing Rossby waves for each case; the dotted and solid coloured lines represent the modenumbers of the barotropic and baroclinic Rossby deformation radii, respectively; and dashed black line represents the modenumber of the topography. For the frequency panels (bottom row): the dashed and solid black lines represent the return and intrinsic frequency of the topography; and the coloured solid lines represent the inertial frequency of the annulus for each case.}
\label{fig:10_FFTs_beta}
\end{figure}

A similar spectra analysis is performed for the experiments that vary $\beta$ (Fig. \ref{fig:10_FFTs_beta}).
The modenumber spectra for all cases with $\beta>1.0$\,rad\,m$^{-1}$\,s$^{-1}$ exhibit peaks at their respective standing Rossby wave modenumber; the energy at these low modenumbers otherwise appears insensitive to $\beta$. 
For high modenumbers, however, increasing the potential vorticity gradient $\beta$ substantially reduces the energy; this reduction is relatively larger for the small $\Delta{T}$ cases.
The large $\Delta{T}$ cases exhibit more energy overall across the modenumber spectra, and the drop-off of energy for high modenumbers is relatively less steep in comparison to the small $\Delta{T}$ cases; the drop-off in all cases appears to steepen past the topographic modenumber.

The frequency spectra exhibit a similar decrease in high frequency energy for increasing $\beta$, especially for the small $\Delta{T}$ cases.
All cases have a maximum at the return frequency of the topography; this spike is relatively larger than its surrounding frequencies for the small $\Delta{T}$ cases.
The low frequency energy for the small $\Delta{T}$ cases is larger than that of the corresponding large $\Delta{T}$ experiments.
For the large $\Delta{T}$ cases, the slope of the frequency spectra appears to steepen past the topographic intrinsic frequency.
In general, the energy at a given frequency for the small $\Delta{T}$ cases is less than that of the respective large $\Delta{T}$ case; the exception to this is at the topography return frequency and the lowest frequencies of the $\beta>1$\,rad\,m$^{-1}$\,s$^{-1}$ cases.
Overall, the effect of increasing the potential vorticity gradient $\beta$ is to reduce the energy at high zonal modenumbers and high frequencies, with this response exacerbated for small $\Delta{T}$ cases.
Increased sidewall temperature difference $\Delta{T}$ tends to increase the variability for almost all modenumbers and frequencies, and especially for modenumbers and frequencies below those associated with the topographic forcing.

\section{Discussion}

\subsection*{Experiment Results}

The experiments clearly demonstrate the distinctly different responses that arise from individually varying the zonal flow speed $U_{b}$, the background gradient of potential vorticity $\beta$, and the sidewall temperature difference $\Delta{T}$.
In terms of the sidewall temperature difference, the dye release visualisations (Fig. \ref{fig:02_dye_release}) are a powerful demonstration of the significant effect that $\Delta{T}$ has on the system.
Reducing $\Delta{T}$ removes a substantial amount of the smaller scale flow features, confines the tracer poleward of the topography, and prevents the tracer from traversing the jet; $\Delta{T}$ has an obvious impact on the overall tracer distribution rate throughout the annulus.
That said, the dye release visualisations also show the remarkable similarities in the larger scale features, in particular the zonal meanders of the jet, which appear relatively insensitive to the sidewall temperature difference $\Delta{T}$.
These qualitative observations are supported by the quantitative analysis; the overall spatial structures of the various temperature fields (Figs. \ref{fig:03_SST_snaps_means_ub}--\ref{fig:06_zonal_means_beta}) and the metrics that represent the waviness of the system (Fig. \ref{fig:07_jet_std}) show these large scale and standing features are insensitive to the sidewall temperature difference $\Delta{T}$.
The small scale and transient features of the system, however, are sensitive to $\Delta{T}$; a reduction in $\Delta{T}$ leads to relatively more standing than transient variability (Fig. \ref{fig:08_variabilities}), and relatively less small scale and high frequency energy (Figs. \ref{fig:09_FFTs_ub}, \ref{fig:10_FFTs_beta}).

The effect of the zonal flow speed $U_{b}$ is most apparent in the zonal lenghtscales of the standing features; these follow what is expected from the canonical relationship described by Equation \ref{eqn:lambda}.
The relative insensitivity of the small scale dynamics to the zonal flow speed is apparent in the surface temperature snapshots shown in Figure \ref{fig:03_SST_snaps_means_ub}; the small scale features here, especially at the peripheries of the annulus, are characteristic of geostrophic turbulence, and of seemingly similar structure across all $U_{b}$ cases.
Indeed, the modenumber spectra appear relatively insensitive to the zonal flow speed, especially for scales smaller than the barotropic Rossby deformation scale.
The waviness of the system appears to follow the zonal length-scale of the standing features, and decreases with decreasing $U_{b}$.
Faster zonal flow speeds do lead to relatively more energy at higher frequencies; however, with the exception of spikes at the topography return frequency, zonal flow speeds have no obvious effect at lower frequencies.

The effect of the potential vorticity gradient $\beta$ is also most apparent in the zonal length-scales of the standing features, and again follows the canonical relationship (Eqn. \ref{eqn:lambda}).
Note that the experiments with $\beta<1.0$\,rad\,m$^{-1}$\,s$^{-1}$ are obviously in a different dynamical regime that is better described by $f$-plane processes; they are useful to include here as they provide insight into the transition from $f$-plane to $\beta$-plane dynamics.
The dominant length-scales of the standing features decrease with increasing $\beta$ as the system transitions from geostrophic to zonostrophic turbulence.
The waviness decreases with increasing $\beta$, which is indicative of a system with relatively stronger radial constraints on its dynamics (Fig. \ref{fig:07_jet_std}b).
As $\beta$ increases, the high modenumber and high frequency energy decreases; this is indicative of an inverse cascade of energy associated with the zonostrophisation of the geostrophic turbulence.

\begin{figure}
\centering
\includegraphics[width=1.0\textwidth]{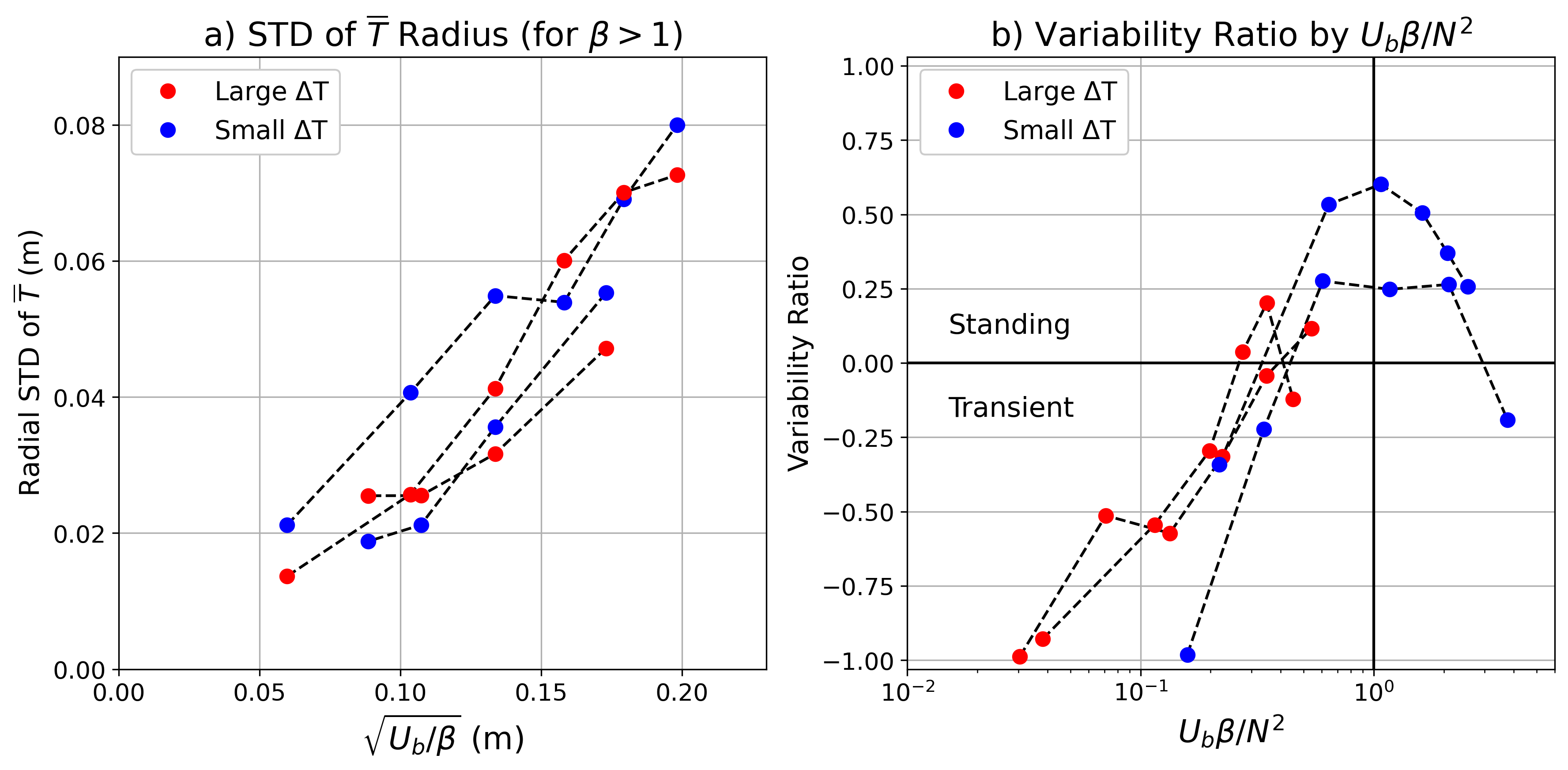}
\caption{Panel a) shows the standard deviation of the radial location of the mean isotherm, a metric which represents the waviness of each case (i.e., Fig. \ref{fig:07_jet_std}), plotted against the square root of the zonal flow speed divided by the potential vorticity gradient, $\sqrt{U_{b}/\beta}$; here we only include experiments with $\beta>1.0$\,rad\,m$^{-1}$\,s$^{-1}$. Panel b) shows the variability ratio (i.e., rightmost column of Fig. \ref{fig:08_variabilities}) plotted against the non-dimensional parameter $U_{b}\beta/N^{2}$.}
\label{fig:11_variability_explained}
\end{figure}

The observation that the waviness of the system increases with zonal flow speed $U_{b}$ and decreases with the potential vorticity gradient $\beta$ is worth exploring further.
If we consider the waviness to be representative of the amplitude of the standing Rossby waves, and that this amplitude is in turn dependent on the wavelength of these standing Rossby waves, then the waviness should behave as per the canonical relationship given in Equation \ref{eqn:lambda}.
Figure \ref{fig:11_variability_explained}a plots the waviness metric shown in Figure \ref{fig:07_jet_std} against the term $\sqrt{U_{b}/\beta}$ for all experiments presented here with $\beta>$1.0\,rad\,m$^{-1}$\,s$^{-1}$.
Expressing the waviness of the experiments with this term collapses data and shows the waviness to increase with $\sqrt{U_{b}/\beta}$.
Note that the sidewall temperature difference $\Delta{T}$ does not appear in the term $\sqrt{U_{b}/\beta}$, and does not appear to influence the waviness of the system.

The nature of the variability appears to be relatively more complicated than the waviness, and seems to depend on all 3 experimental parameters here: $U_{b}$, $\beta$, and the sidewall temperature difference $\Delta{T}$.
The leading effect that the sidewall temperature difference $\Delta{T}$ seems to have is to alter the baroclinicity of the system; the magnitude of $\Delta{T}$ determines whether the topography (forcing) length-scale is larger or smaller than the baroclinic Rossby deformation radius.
We can represent the baroclinicity of the experiment by way of the buoyancy frequency $N$ (Eqn. \ref{eqn:buoyancy}), which is bound by the sidewall temperature difference $\Delta{T}$.
This then allows us to define a non-dimensional number that reflects the 3 experimental parameters here: $U_{b}\beta/N^{2}$.
Note that this term is the only possible non-dimensional combination of the 3 parameters as per the Buckingham-$\pi$ theorem.

Figure \ref{fig:11_variability_explained}b plots the variability ratio by the dimensionless number $U_{b}\beta/N^{2}$ for all 24 experiments.
The variability ratios appear to collapse; the ratios initially increase from predominantly transient for $U_{b}\beta/N^{2}<0.2$, to predominantly standing at $U_{b}\beta/N^{2}\approx0.5$, peaking for $U_{b}\beta/N^{2}\approx1.0$, and decreasing back towards transient variability for further increases in $U_{b}\beta/N^{2}$.
The parameter $U_{b}\beta/N^{2}$ captures the complicated behaviour evident in Figure \ref{fig:08_variabilities} whereby the small and large $\Delta{T}$ experiments exhibited characteristically different responses to changes in $U_{b}$ and $\beta$.

\subsection*{Geophysical Implications}

There is scientific consensus that polar amplification will decrease baroclinicity (i.e., decreased poleward thickness gradient) and reduce the zonal wind speed.
It has been proposed that a subsequent consequence of polar amplification will be the increased north-south meandering of the polar jet stream \cite[e.g.,][]{francis_vavrus2012, francis_etal2017}, although this hypothesis is contested \cite[e.g.,][]{hoskins_woollings2015, blackport_screen2020}.
In our experiments, the waviness or north-south amplitude of the standing Rossby waves appears to be governed by the term $\sqrt{U_{b}/\beta}$, and insensitive to the sidewall temperature difference $\Delta{T}$.
This implies that a reduction in the latitudinal temperature difference alone will not have a direct effect on the north-south amplitude of the standing Rossby waves.
However, the reduction in the latitudinal temperature difference will in turn act to decrease the zonal wind speed; this subsequent reduction in zonal wind speed will then reduce the waviness of the system.
That said, the reduction in the latitudinal temperature difference will act to flatten the latitudinal slope of the tropopause height, which has the dynamical effect of reducing the latitudinal gradient of potential vorticity represented by $\beta$; this would then suggest an increase in the north-south amplitude of standing Rossby waves.
The question then becomes one of sensitivities: for a given change in the latitudinal temperature difference, does the zonal wind speed change relatively more or less than the potential vorticity gradient?
We hypothesise that the zonal wind speed is relatively more sensitive to the latitudinal temperature difference than the potential vorticity gradient, but acknowledge that further insight is needed.

Another suggested consequence of polar amplification is the enhanced prevalence of prolonged and extreme mid-latitude weather events such as drought, flooding, heat waves, and cold spells.
In our experiments, the duration of conditions is reflected by the partitioning of the total variability into its standing and transient components (the variability ratio; rightmost column of Fig. \ref{fig:08_variabilities}), where the transient variability represents high-frequency and/or small-scale storms, and the standing variability represents the more persistent and extreme weather events.
The response of the variability ratio is sensitive to all 3 experimental parameters.
Reducing the sidewall temperature difference $\Delta{T}$ leads to a greater prevalence of standing events (i.e., more prolonged and extreme weather events).
Reducing the zonal flow speed $U_{b}$ and/or the potential vorticity gradient $\beta$ can increase the variability ratio for small $\Delta{T}$ while decreasing it for large $\Delta{T}$.
Casting the results in terms of the non-dimensional parameter $U_{b}\beta/N^{2}$ collapses these behaviours, and appears to suggest a maximum in the standing variability partition for $U_{b}\beta/N^{2}\approx1$.
For systems with $U_{b}\beta/N^{2}<1$, a reduction in $U_{b}\beta/N^{2}$ (i.e., a reduction in $U_{b}$ and/or $\beta$, or an increase in $N^{2}$) will tend to shift the variability towards more transient conditions, while for $U_{b}\beta/N^{2}>1$, a reduction in $U_{b}\beta/N^{2}$ tends to increase the standing component.

We can estimate the order of magnitude of $U_{b}\beta/N^{2}$ for the mid-latitudes; if we consider $U_{b}\sim10^{1}$--$10^{2}$\,m\,s$^{-1}$, $\beta\sim10^{-10}$\,rad\,m$^{-1}$\,s$^{-1}$, and $N\sim10^{-4}$--$10^{-2}$\,s$^{-1}$, the term $U_{b}\beta/N^{2}$ has an upper bound of 1 (i.e., using the largest value of $U_{b}$ and smallest value of $N$).
This implies that a reduction in zonal wind speed $U_{b}$ will shift the system to more transient events, while a reduction in the stratification or baroclinicity $N$ with shift the system towards more standing events.
Predicting how the mid-latitudes will respond to polar amplification then requires quantifying the relative sensitivities of $U_{b}$ and $N$ to the latitudinal temperature difference $\Delta{T}$.

\section{Conclusions}

The experiments here focus on understanding the dynamical consequences of polar amplification.
The seemingly simple act of reducing the sidewall temperature difference $\Delta{T}$ in an idealised annulus experiment excites a global circulation response; small-scale phenomena are virtually eliminated, and the distribution of passive tracer throughout the tank is reduced.
The larger scale dynamics of the system, however, are remarkably similar; these robust features appear surprisingly insensitive to the temperature difference.
Subsequent quantitative analysis with thermal camera measurements confirm these features to be standing Rossby waves, and their responses to changes in the zonal flow speeds $U_{b}$ and background potential vorticity gradient $\beta$ are captured by established canonical Rossby wave theory; their north-south amplitude and zonal wavelength are not directly sensitive to the sidewall temperature difference, only $U_{b}$ and $\beta$, as $\sqrt{U_{b}/\beta}$.

The nature of the variability in the system, however, is sensitive to the sidewall temperature gradient $\Delta{T}$.
A large $\Delta{T}$ exhibits relatively more energy at high modenumbers and high frequencies in the system, which is characteristic of a more baroclinic state supported by a larger stratification $N$.
Reducing $\Delta{T}$ tends to shift variability from time-dependent transient phenomena towards time-invariant standing features.
The variability response to the experiment parameters is well described by the non-dimensional term $U_{b}\beta/N^{2}$.

The idealised experiments presented here offer valuable insight into why the dynamical consequences of polar amplification remain an open question.
The ability to prescribe the experimental parameters independently of one another enables a thorough investigation into the sensitivities of the system.
In the real world, these parameters are not able to be decoupled; the reduction in latitudinal temperature difference associated with polar amplification will affect the zonal wind speed, potential vorticity gradient, and mid-latitude stratification all at different rates, with these rates dependent on feedbacks from the other parameters.
That said, the idealised laboratory experiments with the LRA continue to provide great insight into Rossby waves and mid-latitude dynamics; future experiments will explore the seasonality of the system and dynamical constraints on latitudinal heat fluxes.

\clearpage
\section*{Acknowledgments}

We wish to thank Angus Rummery, Peter Lanc, and Tony Beasley for the construction of the apparatus, technical support, and laboratory assistance.
We acknowledge very helpful discussions with Marty Singh, Michael Barnes, Julie Arblaster, Deepashree Dutta, Martin Jucker, and Navid Constantinou, and are indebted to the academic brilliance of Ross Griffiths.
TGS was supported by an {\it{Australian Research Council Centre of Excellence for Climate Extremes}} Undergraduate Research Scholarship.
The images in Figure \ref{fig:02_dye_release} are available as a video on the {\it{FluidsIn4K}} youtube channel at {\it{www.youtube.com/watch?v=CatIj6DU6ss}}
Upon acceptance, the surface temperature field measured by the thermal camera will be published at Zenodo and the digital object identifier (doi) will be quoted here.


\bibliography{stewart_etal_JCli_bib.bib}
\bibliographystyle{gGAF}

\
\end{document}